\documentclass[prb,twocolumn,groupeaddress]{revtex4-2}

\usepackage{amsmath}
\usepackage{graphicx}
\usepackage[colorlinks=true, allcolors=blue]{hyperref}

\usepackage{lineno}
\usepackage{ulem}

\begin{document}

\title{Electronic correlations and superconducting instability in La$_3$Ni$_2$O$_7$ under high pressure}
\author{Frank Lechermann}
\affiliation{Institut f\"ur Theoretische Physik III, Ruhr-Universit\"at Bochum,
  D-44780 Bochum, Germany}
  \author{Jannik Gondolf}
\affiliation{Institut f\"ur Theoretische Physik III, Ruhr-Universit\"at Bochum,
  D-44780 Bochum, Germany}
  \author{Steffen B\"otzel}
  \affiliation{Institut f\"ur Theoretische Physik III, Ruhr-Universit\"at Bochum,
  D-44780 Bochum, Germany}
\author{Ilya M. Eremin}
\affiliation{Institut f\"ur Theoretische Physik III, Ruhr-Universit\"at Bochum,
  D-44780 Bochum, Germany}

\begin{abstract}
 Motivated by the report of superconductivity in bilayer La$_3$Ni$_2$O$_7$ at high pressure, we examine the interacting electrons in this system.
 First-principles many-body theory is utilized to study the normal-state electronic properties. Below
 100\,K, a multi-orbital non-Fermi liquid state resulting from loss of Ni-ligand coherence within a flat-band dominated low-energy landscape is uncovered. The incoherent low-temperature Fermi surface displays strong mixing between Ni-$d_{z^2}$ and Ni-$d_{x^2-y^2}$ orbital character. In a model-Hamiltonian picture, spin fluctuations originating mostly from the Ni-$d_{z^2}$ orbital give rise to strong tendencies towards a 
 superconducting instability with $B_{1g}$ or $B_{2g}$ order parameter. 
 The dramatic enhancement of $T_{\rm c}$ in pressurized La$_3$Ni$_2$O$_7$ is due to stronger Ni-$d_{z^2}$ correlations compared to those in the infinite-layer nickelates.
\end{abstract}

\maketitle

\textit{Introduction.---}
In a recent finding, Sun {\sl et al.}~\cite{sun23} reported superconductivity near a temperature $T=80$\,K in bulk single crystalline La$_3$Ni$_2$O$_7$ at pressures $p>14$\,GPa. This adds a whole new chapter to the young research field of superconducting (SC) nickelates, hosting high-$T_{\rm c}$ cuprate-akin NiO$_2$ square-lattice planes. The field has been inaugurated by the discovery of electron pairing in thin films of Sr-doped infinite-layer NdNiO$_2$ with a $T_{\rm c}\sim 15$\,K in 2019~\cite{li19}. Early follow-up studies~\cite{li19,li20,zeng20,osada20,osada21,pan21,zeng22} detected similar SC phases in thin films of Pr$_{1-x}$Sr$_x$NiO$_2$, La$_{1-x}$Sr$_x$NiO$_2$ and also in multilayer Nd$_6$Ni$_5$O$_{12}$ thin films. While these reduced 
SC nickelates share a common motif by the lack of apical oxygens (resulting from a topotactic reaction) and Ni$(3d^{9-\delta})$ oxidation states, the characteristics of bilayer La$_3$Ni$_2$O$_7$ differ. It still holds the apical oxygens and Ni formally has $3d^{7.5}$ configuration. Furthermore, a comparison to high-$T_{\rm c}$ cuprates with their key Cu-$d_{x^2-y^2}$ single-orbital character becomes quite stretched. Whereas there is an ongoing debate concerning a dominant single Ni-$d_{x^2-y^2}$~\cite{wu19,zhangzhang20,karp20,leonov20,adhikary20,Kitatani2020,been21,geisler21,gu2020substantial,plienbumrung22,jiang22,Held2022} versus dominant multi-orbital Ni-$e_g$~\cite{werner20,lechermann20-1,lechermann2020multiorbital,petocchi20,kang21} low-energy physics in reduced SC nickelates, the nominal hole doping away from Ni$(3d^9)$ is that large in La$_3$Ni$_2$O$_7$ as to render Ni multi-orbital physics inevitable. In this context, a Ni-$e_g$ multi-orbital picture for infinite-layer nickelates results in a competition between SC instabilitites of varying flavor~\cite{kreisel22}.

On a wider scope, two further issues appear relevant.
First, even if the formal oxidation state reads accordingly, a Ni$(3d)$ occupation well below $n_d=8$ is hardly occuring in known nickel oxides. Instead in most cases, a $3d^8L$ state incorporating holes on ligand oxygen is realized~\cite{dem93,miz00,par12,joh14,sub15,bis16,lechermann22-2}, also accompanied by a smaller charge-transfer energy $\Delta=\varepsilon_d-\varepsilon_p$ between Ni$(3d)$ and O$(2p)$. Secondly, bilayer oxides from the $p$-layered Ruddlesden-Popper series A$_{p+1}$TM$_p$O$_{3p+1}$ (A: rare-earth, alkaline-earth; TM: transition; metal) often display a much more delicate normal-state low-energy physics than corresponding single-laver systems. This is e.g. exemplified for ruthenates~\cite{perry01,borzi07,lechermann-327} and iridates~\cite{king13}.
Previous theoretical accounts of bilayer lanthanum nickelate focussed on the paramagnetic metal~\cite{zhang94,taniguchi95,wu01} at ambient pressure. From density-functional theory (DFT), a charge-density wave state was predicted~\cite{seo96}, while DFT+Hubbard $U$ considerations~\cite{pardo11} remark the possible relevance of magnetically-ordered states. 

In this work, we provide a theoretical description of the correlated electronic structure of paramagnetic La$_3$Ni$_2$O$_7$ under high pressure, by employing the combination of DFT, self-interaction correction (SIC) and dynamical mean-field theory (DMFT), i.e. the so-called DFT+sicDMFT approach~\cite{lechermann19}. Moreover, a model Hamiltonian perspective via the random-phase approximation (RPA) onto the possible superconducting instabilitites from spin fluctuations is presented. We reveal an intriguing low-energy physics of pressurized La$_3$Ni$_2$O$_7$ in the normal state. The highly-correlated interplay between Ni-$d_{z^2}$, Ni-$d_{x^2-y^2}$ and O$(2p)$, with the former displaying partial flat-band character, gives rise to a distinct non-Fermi-liquid (NFL) regime below $T<100$\,K. The model-RPA investigation points to a 
$B_{1g}$ or $B_{2g}$ SC order parameter that would emerge from this multi-orbital scenario. Most importantly, we argue that the Ni-$d_{z^2}$ orbital is much more correlated in the bilayer case than in infinite-layer nickelates. Consequently, it interacts in a much more concerted fashion with Ni-$d_{x^2-y^2}$, which could explain the much higher $T_c$ in the bilayer case. Note that Ni-$d_{x^2-y^2}$ is always close to half filling in superconducting nickelates and alone cannot explain the difference in $T_{\rm c}$ between these two classes of materials.

\textit{Theoretical Approach.---}
The charge self-consistent~\cite{grieger12} DFT+sicDMFT framework~\cite{lechermann19}, where the Ni sites act as 
DMFT impurities and Coulomb interactions on oxygen enter 
by SIC on the pseudopotential level~\cite{korner10}, is put into practise. The DFT part consists of a mixed-basis pseudopotential code~\cite{elsaesser90,lechermann02,mbpp_code} and SIC is applied to the O$(2s,2p)$ orbitals via weight factors $w_p$. While the $2s$ orbital is fully corrected with $w_p=1.0$, the choice~\cite{korner10,lechermann19,lechermann20-1} $w_p=0.8$ is used for $2p$ orbitals. Continuous-time quantum Monte Carlo in hybridization expansion~\cite{werner06} as implemented
in the TRIQS code~\cite{parcollet15,seth16} solves the DMFT problem. A five-orbital general Slater-Hamiltonian, parameterized by Hubbard $U=10$\,eV and Hund exchange $J_{\rm H}=1$\,eV
\cite{lechermann20-1,lechermann2020multiorbital}, governs the correlated subspace defined by Ni projected-local orbitals~\cite{amadon08}. Crystallographic data are taken from experiment~\cite{sun23}. 
Further calculational details are given in the supplementary material (with Refs.~\cite{kotliar06, lechermann21,chen22,anisimov93,luo23,jar96,vid77,interparc}).

\textit{Results.---}
\begin{figure}[t]
      \includegraphics[width=\linewidth]{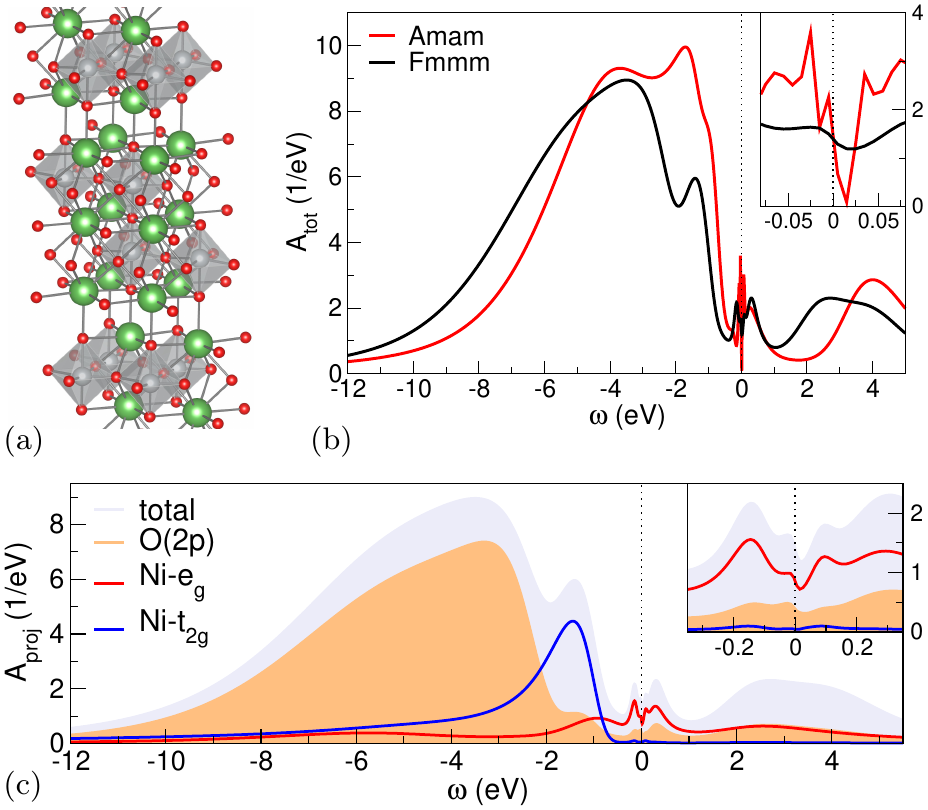}
	\caption{{\bf k}-integrated electronic spectrum from DFT+sicDMFT at $T=80$\,K.
  (a) $Fmmm$ crystal structure of La$_3$Ni$_2$O$_7$ with $c$-axis vertical: La (green), Ni(grey) and O(red).
  Note the bilayer of edge-sharing NiO$_6$ octahedra in the centre and along $c$, bounded by  LaO$_2$ layers up and below.
  (b) Total spectral function of low-pressure $Amam$ and high-pressure $Fmmm$ phase (inset: low-energy blow up). (c) Site- and orbital-projected spectral function for the $Fmmm$ phase (inset: low-energy blow up).} \label{fig1}
\end{figure}	
In experiment, there is a structural transition in La$_3$Ni$_2$O$_7$ from a low-pressure $Amam$ phase with finite NiO$_6$ octahedral tilting to a high-pressure $Fmmm$ phase without tilting~\cite{sun23}. The $Amam$ (space group 63) crystal structure has four equivalent Ni sites in the primitive cell, in contrast to two Ni sites for the $Fmmm$ (space group 69) one.
Key feature of the latter structure at $p=29.5$\,GPa (see Fig.~\ref{fig1}a) is a rather small distance of $\sim 1.76$\,\AA\, between Ni and apical oxygen within the bilayer. The calculations show (cf. Fig.~\ref{fig1}b) that the spectrum of the $Amam$ phase 
at $p=1.6$\,GPa is (nearly) gapped, in line with the measured different transport properties~\cite{sun23}.
For the rest of the paper, we will restrict the discussion to the properties of the high-pressure $Fmmm$ phase.

The electronic spectrum in Fig.~\ref{fig1}c exhibits a metallic state with Ni-$e_g\,\{d_{z^2}, d_{x^2-y^2}\}$ and O$(2p)$ character at the Fermi level $\varepsilon_{\rm F}$, but lacks strong quasiparticle (QP) signature. The Ni-$t_{2g}\,\{d_{xz}, d_{yz}, d_{xy}\}$ manifold is mostly filled with a peak at $\sim -1.5$\,eV. The main O$(2p)$ weight peaks at $\sim -3.5$\,eV. Sizable O$(2p)$ weight in the unoccupied part of the spectrum points to ligand-hole states. And indeed, the integrated projected spectral parts yield occupations $n_d=7.98$ and $n_p=5.60$, resulting in a substantial content of 0.4\,holes per oxygen and a near Ni$(3d^8)$ configuration. Thus maybe nonsurprisingly for this high-pressure system, the degree of covalency is significant and about 1.8 formula-unit-cell valence electrons have to reside in states of La$(5d6s)$ and/or Ni$(4s)$ character. The DFT+sic computed charge transfer energy $\Delta=3.6$\,eV lies inbetween the infinite-layer values of 5.0\,eV for NdNiO$_2$ and of 1.3\,eV for SrCuO$_2$~\cite{lechermann20-1}. Note in this respect that while the Ni-$e_g$ character dominates at $\varepsilon_{\rm F}$, the corresponding O$(2p)$ content is still larger than in NdNiO$_2$, hinting to a comparatively enhanced role of oxygen degrees of freedom at low-energy.
\begin{figure}[b]
      \includegraphics[width=\linewidth]{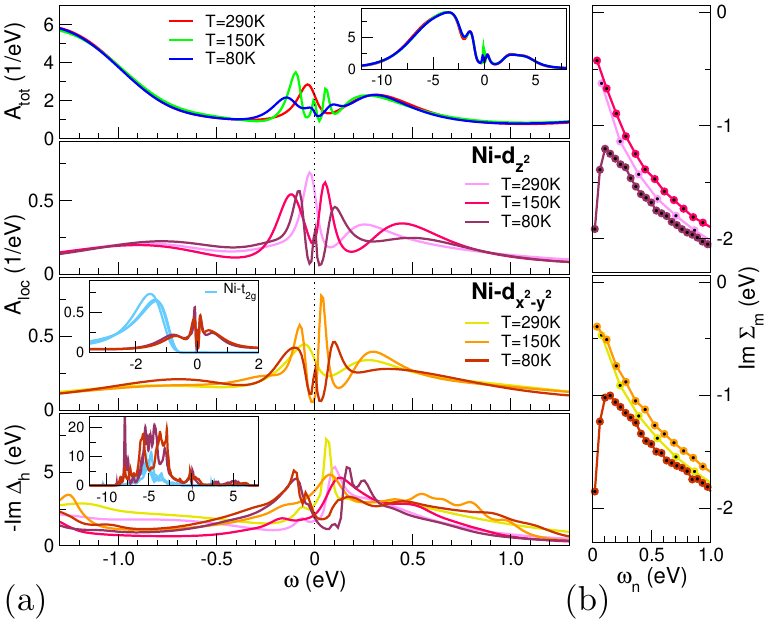}
	\caption{FL-like vs. NFL behavior with $T$ in the $Fmmm$ phase from DFT+sicDMFT. (a) Top panel: total spectral function (inset: wide energy scale). Middle panels: local Ni-$e_g$ spectral function (inset: wide energy scale including local Ni-$t_{2g}$ spectra for $T=80$\,K). Bottom panel: Ni-$e_g$ hybridization function (inset: same protocol as for middle panels). (b) Imaginary part of the Ni-$e_g$ self-energies $\Sigma(i\omega_n)$ on the Matsubara axis (colors according to middle panel of (a)). } \label{fig2}
\end{figure}	
\begin{figure*}[t]
		\includegraphics[width=\linewidth]{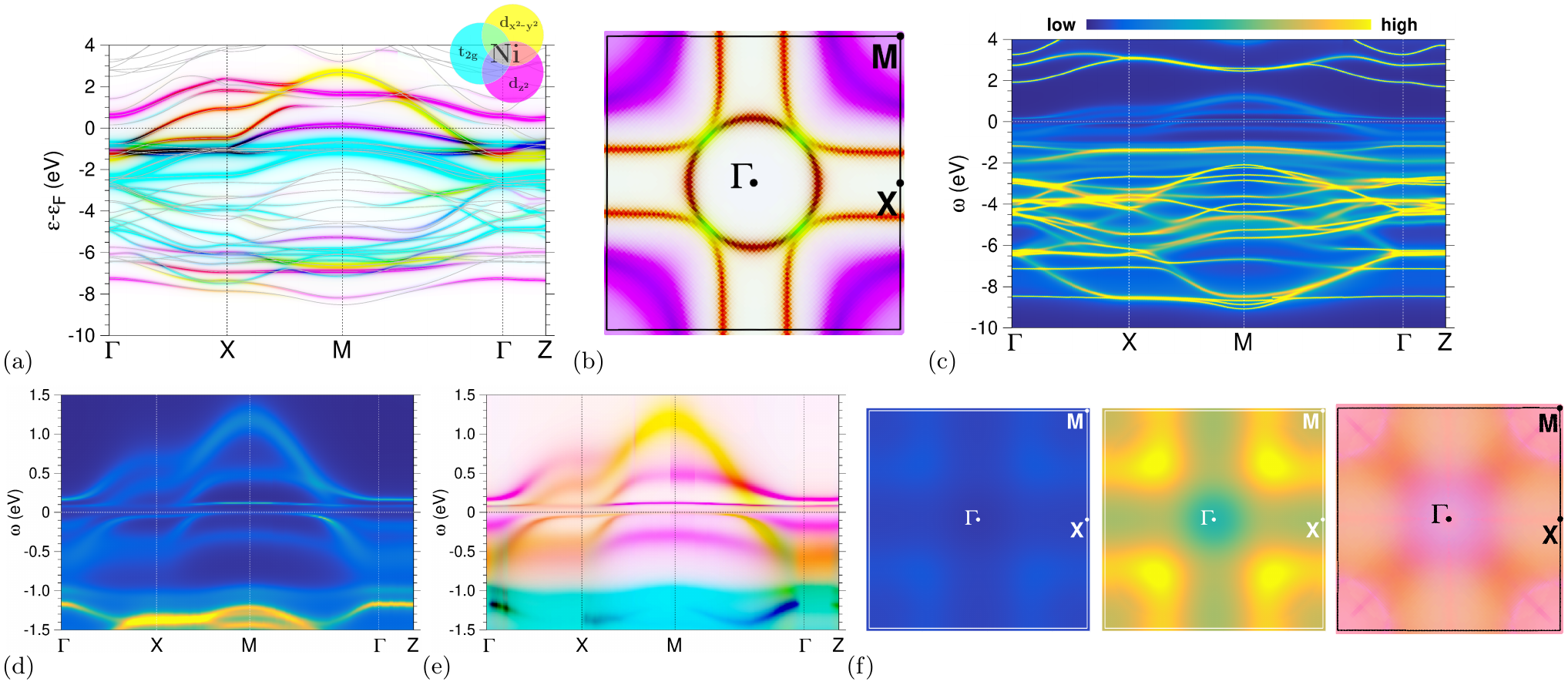}
	\caption{{\bf k}-dependent spectral features of the $Fmmm$ phase from DFT (a,b) and DFT+sicDMFT at $T=80$\,K (c-f).
 (a) DFT band structure in Ni$(3d)$-fatspec representation. Note that red, green and purple colors correspond to mixed-orbital states according to the color-coding inset. (b) DFT $k_z=0$ Fermi surface in Ni-fatspec representation. (c) Interacting spectral function in large energy window, and (d) in small window around $\varepsilon_{\rm F}$. (e) Same as (d) but in Ni-fatspec representation. (f) Interacting $k_z=0$ Fermi surface from left to right: regular intensity, 10$\times$-increased intensity, Ni-fatspec representation. } \label{fig3}
\end{figure*}

As anticipated for a bilayer oxide, and already documented by the subtle $Amam$ vs. $Fmmm$ low-energy difference, the electronic states near the Fermi level are delicate. To reach a better understanding of the relevant coherence scales, we therefore performed additional calculations at higher $T$. Figure~\ref{fig2} shows that the low-energy regime reacts sensitively to temperature. The total spectral function (top panel of Fig.~\ref{fig2}a) evolves from a QP-like structure at room temperature to a flattened weight around the Fermi level at $T=80$\,K. On the local level (middle panel of Fig.~\ref{fig2}a), it is first noted that Ni-$d_{z^2}$ and Ni-$d_{x^2-y^2}$ are both half filled. Second, upon lowering $T$, the near-$\varepsilon_{\rm F}$ $A_{\rm loc}(\omega)$ transforms from QP-like, to pseudogap and finally to low-amplitude peak. A link between total and local spectrum may be established via the hybridization function $-{\rm Im}\Delta_{\rm h}$, shown in bottom panel of Fig.~\ref{fig2}a. It displays a (pseudo)gap at $T=80$\,K, altogether rendering it obvious that a NFL state is reached. The Ni-$e_g$ self-energies shown in Fig.~\ref{fig2}b underline this picture with a low-frequency upturn at low $T$. A Fermi-liquid (FL) fit to the room-$T$ data yields effective masses $m^*_{z^2}=6.4$ and $m^*_{x^2-y^2}=5.6$. But note that even the ambient system is not a good FL. Though a linear-in-frequency regime holds for smallest Matsubara $\omega_n$, the scattering rate
$\sim -\lim_{\omega_n\rightarrow 0}\Sigma(i\omega_n)$
remains quite large for the given $T$. And already the $T=150$\,K data displays further NFL tendencies. Because of half-filled Ni-$e_g$ (one electron in each of the two orbitals) as well as the very low-energy scale for non-QP formation, a sole $U$- or $J_{\rm H}$-driven NFL behavior seems not likely.

Let us thus turn to the {\bf k}-resolved spectra to gain insight into the origin of the NFL behavior. Figure~\ref{fig3}(a,b) displays the DFT band structure and Fermi surface to set the stage. The Ni fatspec representation~\cite{lechermann22}
marks the dominant Ni$(3d)$ character, showing that there are majorly four Ni-$e_g$ dispersions, associated with the two equivalent Ni sites in the unit cell, governing the low-energy region. The inner two bands form a ($\alpha$) electron pocket around $\Gamma$ and a ($\beta$) hole sheet opening towards $X$. Note that those two Fermi-surface sheets are strongly Ni-$d_{z^2}/d_{x^2-y^2}$ mixed. The upper (anti-bonding) Ni-$d_{z^2}$-dominated band, also having sizable apical O$(2p)$ character, is not crossing $\varepsilon_{\rm F}$. Instead, a self-doping mainly La-based band mingles into the Ni-$e_g$ fourfold dispersion and gives rise to a large electron pocket around $Z$. Finally, the lower (bonding) Ni-$d_{z^2}$-dominated band forms flattened ($\gamma$) hole pockets around $M$. The Ni-$t_{2g}$ character very weakly mixes into part of the Fermi-surface sheets, but otherwise has major weight below $\varepsilon_{\rm F}$ and does not play a key role for the low-energy physics.

With correlations at $T=80$\,K, i.e. deep into the NFL regime, the low-energy picture changes quite dramatically (see Fig.~\ref{fig3}c-f). First, the near-$\varepsilon_{\rm F}$ dispersions become generally very weak in intensity. Figure~\ref{fig3}d shows that while the dispersions {\sl away} from the Fermi level still keep some renormalized coherence, {\sl right at} $\varepsilon_{\rm F}$ they dissolve. This is orthogonal to the understanding of a FL state and marks the strong NFL nature of the pressurized bilayer system. Accordingly, the (weakly $k_z$-dependent) interacting Fermi surface displayed in Fig.~\ref{fig3}f becomes very weak and blurry. Only when raising the representation intensity (middle part of Fig.~\ref{fig3}f), a hole-like "sheet" structure around $M$ emerges. It is reminiscent of the original DFT Ni-$d_{z^2}$ flat-band-based $\gamma$ sheet, but with stronger mixed Ni-$e_g$ character (cf. right part of Fig.~\ref{fig3}f)). The enhanced intermixing presumably comes from a correlation-induced meet up with the Ni-$d_{x^2-y^2}$ branded $\beta$ sheet in $\Gamma$-$M$ direction. This is also supported from the disappearance of the self-doping $Z$ pocket from the Fermi surface. Such strong correlation-induced shifts of self-doping bands have already been observed in other nickelates~\cite{lechermann22,worm22,lechermann22-2}.
It becomes intuitively obvious that all these very-low-energy ramifications in the interacting regime have to strongly build up on the present flat-band scenario. There, the introduced renormalizations create a large phase space for intriguing quantum fluctuations, rendering robust QP formation 
impossible~\cite{sayyad20}.

Albeit the NFL character may be relevant for superconductivity, let us get a first handle on SC instabilities from a weak-coupling perspective for coherent Fermi-surface sheets and leave the discussion of the role of the NFL behavior and its relevance for superconductivity to future studies. To do this, we constructed a $4\times 4$ maximally-localized Wannier~\cite{marzari12} Hamiltonian for the Ni-$e_g$ based DFT bands. It carries hopping integrals $t_{ij}^{\ell \ell'}$ for $\ell,\ell'=1d_{z^2},1d_{x^2-y^2},2d_{z^2},2d_{x^2-y^2}$ and lattice sites $i,j$. Here $1$ and $2$ refer to the two Ni sublattices in the $Fmmm$ structure.
\begin{figure}[t]
      \includegraphics[width=\linewidth]{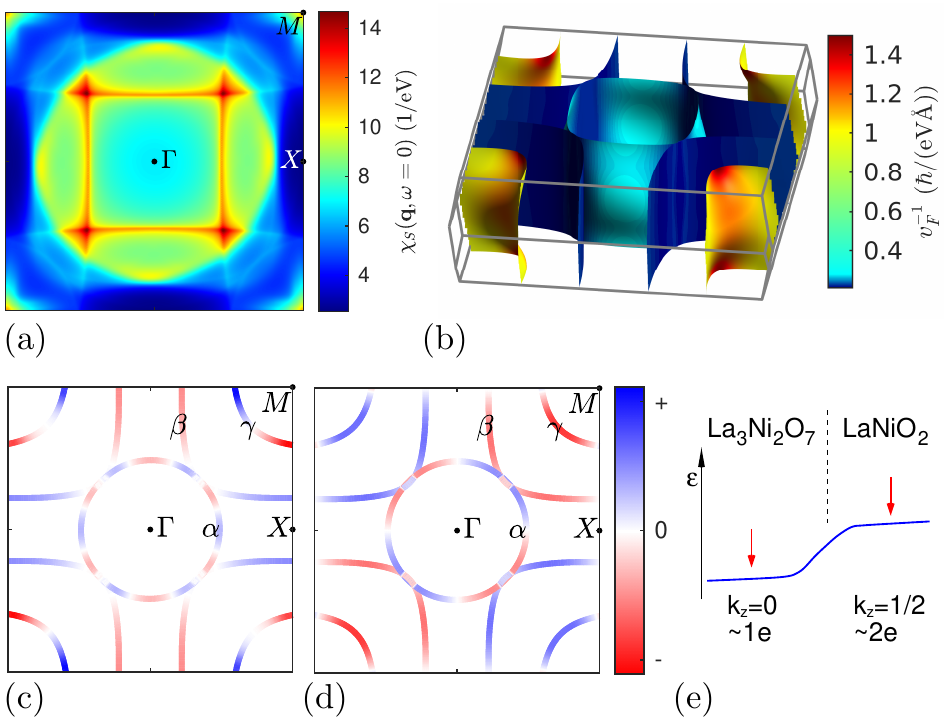}
	\caption{(a) RPA spin susceptibility $\chi_\mathrm{S}(\bf{q}, \omega=0)$ for $\bar{U}=0.36$ and $\bar{J}_{\rm H}=\bar{U}/7$ at $T=80$\,K. (b) Absolute value of the inverse Fermi-surface velocity, showing significant anisotropy. (c,d) Leading $B_{1g}$ and $B_{2g}$ solutions of the linearized gap equation for same $\bar{U}$ and $T$ as in (a) and $\bar{J}_{\rm H}=\bar{U}/4$ and $\bar{J}_{\rm H}=\bar{U}/7$, respectively. (e) Sketch for hole doping (red arrows) near the SC regime, relating flat-band scenarios of pressurized bilayer and reduced nickelates (see text).} \label{fig4}
\end{figure}	
 Adding local Coulomb interactions, the effective Hamiltonian reads 
\begin{equation}                                   
H=\sum_{i\ne j,\ell \ell',\sigma} t_{ij}^{\ell \ell'}
\,c^\dagger_{i\ell\sigma} c^{\hfill}_{j\ell'\sigma}+\sum_i\left( H^{(i)}_{\rm int} + H^{(i)}_{\rm orb}\right),\label{eq_ham}   
\end{equation}
The onsite interaction $H^{(i)}_{\rm int}$ has Slater-Kanamori form, i.e. includes density-density terms as well as pair-hopping and spin-flip terms, parameterized by $\Bar{U}$ and $\Bar{J}_{\rm H}$ (here given in units of the tight-binding bandwidth). Note that within the downfolded model, the Hubbard interaction is stronger screened, resulting in a smaller value than in the comprehensive DFT+sicDMFT treatment. The remaining non-interacting onsite $H^{(i)}_{\rm orb}$ carries crystal-field terms via onsite levels $\varepsilon_\ell$.
In order to investigate the pairing symmetry starting from the effective band structure, we employ the standard multiorbital RPA treatment, developed by Graser {\sl et al.}~\cite{graser2009near}, to derive a linearized gap equation. This treatment is i.e. a pertubative weak-coupling expansion in the Hubbard-Kanamori-type interaction and provides a reliable first insight into the leading pairing instabilities in layered unconventional superconductors, including nickelates~\cite{kreisel22}. Further details are presented in the supplementary material.
We calculate the RPA spin (charge) susceptibility $\chi_S(\mathbf{q},\omega=0)$ $(\chi_C)$ based on a two-dimensional cut through the first and second Brillouin zone to include relevant scatterings between the $k_z=0$ and $k_z=1/2$ layers. Figure~\ref{fig4}a shows the obtained peak structure at $T = 80$\,K, which is similar to previous findings~\cite{luo23}. The inverse of the band velocities $\mathbf{v}^{\mu}_F = \hbar^{-1} \mathbf{\nabla}_{\mathbf{k}}\epsilon^{\mu}_{\mathbf{k}}$ at the Fermi surface is illustrated in Fig.~\ref{fig4}b. Here $\epsilon^{\mu}_{\mathbf{k}}$ is the $\mu$-th eigenvalue of the non-interacting Hamiltonian. The anisotropy of the band velocities proves relevant when solving the linearized gap equation.  
For $\bar{J}_{\rm H}=\bar{U}/4$ the $B_{1g}$ solution depicted in Fig.~\ref{fig4}c is the leading solution. When the Hund exchange is weaker, the $B_{2g}$ solution is leading. It is shown in Fig.~\ref{fig4}d for $\bar{J}_{\rm H}=\bar{U}/7$. A sub-leading $s_{\pm}$-wave solution, which belongs to the $A_{1g}$ irreducible representation, becomes dominant when alterations to the band structure either reduce the anisotropy of the inverse band velocity on the $z^2$-dominated $\gamma$-sheet near the $M$-point, and/or when the van-Hove singularity at the $X$ shifts closer to the Fermi surface. This shift effectively also increases the inverse of the Fermi velocity of the $\beta$ sheet. A more detailed discussion of the distinct solutions and their dependency on the details of the band structure is presented in the supplementary material, using the tight-binding model by Luo {\sl et al.}~\cite{luo23} as an additional illustration.
The total pairing strength given by the leading eigenvalue is strongly driven by $\bar{U}$, and generally, superconductivity would sensitively react to the level of of Ni-$d_{x^2-y^2}$/$d_{z^2}$ incoherence. This issue should be a subject of further theoretical and experimental study.

\textit{Discussion.---}
We have shown that the peculiar correlated electronic structure 
and concomitant SC instability of La$_3$Ni$_2$O$_7$ originates from the interplay of half-filled Ni-$e_g$ orbitals within a Ni-$d_{z^2}$-created flat-band scenario. The role of O$(2p)$ is enhanced compared to reduced nickelates, yet a decisive role cannot be deduced from this initial theory study. But note that the Ni-O distance along $c$ within the bilayer turns out remarkably small, thus inter-site Ni-Ni self-energies may not be negligible. Those could e.g. be addressed in a two-site, two-orbital cluster-DMFT study, which however is beyond the present scope. Comparing to the phenomenology of reduced nickelates, a line can be drawn between these and the bilayer system as sketched in Fig.~\ref{fig4}e for a modelized single-Ni unit-cell system: In the reduced systems, the hole doping relevant for superconductivity occurs mainly in the Ni-$d_{z^2}$ upper-branch flat-band part around  $k_z=1/2$~\cite{lechermann2020multiorbital,lechermann21}, corresponding to a well-filled Ni-$d_{z^2}$ orbital. For pressurized La$_3$Ni$_2$O$_7$ however, the hole doping takes place in the lower-branch flat-band part around $k_z=0$. There, Ni-$d_{z^2}$ is close to half-filling, much more correlated and therefore more on par with Ni-$d_{x^2-y^2}$. This should be the reason for the different $T_{\rm c}$ in the unlike nickelates. One may speculate that this different flat-band doping regime can also be realized in reduced multilayer nickelates~\cite{zha17,pan21,lechermann22} via tailored doping protocols even at ambient pressure.

\section{Acknowledgements} 
The work is supported by the German Research Foundation within the bilateral NSFC-DFG Project ER 463/14-1. 
Computations were performed at the Ruhr-University Bochum and the JUWELS Cluster of the J\"ulich Supercomputing Centre (JSC) under project miqs.

\bibliography{literatur}

\clearpage
{\bf Supplementary Material}

\subsection{DFT+sicDMFT calculations and presentation}
In the following, details of the DFT+sicDMFT settings as well as of the presentation of the corresponding results are provided. For a background on the general theoretical framework we refer to Ref.~\cite{kotliar06} for DFT+DMFT and to Ref.~\cite{lechermann19,lechermann21,chen22} for DFT+sicDMFT for nickelates.
For the DFT part, a mixed-basis pseudopotential framework in the local-density approximation (LDA) is put into practice. A $9\times9\times 9$ for $Fmmm$ and a $5\times5\times 5$ $k$-point mesh for $Amam$ is utilized and the plane-wave cutoff energy is generally set to $E_{\rm cut} =16$\,Ry. Local basis orbitals are introduced for La$(5d)$, Ni$(3d)$ and O$(2s, 2p)$. The role of possible spin-orbit effects is neglected in the crystal calculations, but the pseudopotentials are generated with including spin-orbit coupling.
The DMFT correlated subspace on each Ni site is governed by a rotational-invariant five-orbital Slater Hamiltonian. It is applied to the Ni$(3d)$ projected-local orbitals. The projection
is performed on the Kohn-Sham bands above the 7(14) bands of dominant O$(2s)$ character for $Fmmm$($Amam$). The projection window spans 34(68) bands, including the KS states of dominant Ni$(3d)$ and O$(2p)$, as well as one additional KS band for each La site in the primitive cell. A Hubbard $U=10$\,eV and a Hund exchange $J_{\rm H}=1.0$\,eV are chosen to parametrize the local interacting Hamiltonian. No explicit Coulomb interactions beyond DFT are introduced on La. The fully-localized-limit double-counting scheme ~\cite{anisimov93} is applied. Continuous-time quantum Monte Carlo in hybridization expansion as implemented in the TRIQS code is used to solve the multisite DMFT problem. Up to $1.5\cdot 10^9$ Monte-Carlo sweeps are performed to
reach convergence. A Matsubara mesh of 1025(2049) frequencies is used to account for the higher(lower)-temperature regime. In detail, this means that 1025 frequencies are used for $T>100$\,K and 2049 frequencies are utilized for $T<100$\,K. For the analytical continuation from Matsubara space onto the real-frequency axis, the Maximum-entropy method~\cite{jar96} is used for the {\bf k}-integrated spectra (by continuation of the Bloch Green’s function) and the Pad{\'e} method~\cite{vid77} is employed
for the {\bf k}-resolved spectra (by continuation of the local self-energies). 

Note that in the present work, the DMFT hybridization function, reading $\Delta_{\rm hyb}(\omega)=\omega+\mu-G_{\rm loc}^{-1}(\omega)-\Sigma_{\rm imp}(\omega)$, is constructed from using the Pad{\'e}-method continued local self-energies. Three different versions of the {\bf k}-integrated spectral function $A(\omega)$ are discussed in the main text. The local spectral function $A_{\rm loc}$ of the Ni$(3d)$ orbitals is directly constructed from the local DMFT Greens function $G_{\rm loc}$, whereas the projected spectral function $A_{\rm proj}$ is derived from the interacting Bloch Greens function $G_{\rm bl}$ of the full DFT+(sic)DMFT Hilbert space~\cite{amadon08,grieger12}. Thus the projected spectral function covers the full hybridization effects in the interacting lattice regime, whereas the local spectral function focuses on the correlated subspace. The total spectral function $A_{\rm tot}$ represents the sum of the site- and orbital-resolved $A_{\rm proj}$. In that sense, $A_{\rm tot}$ and $A_{\rm proj}$ are the interacting counterparts of the total as well as site-and orbital-resolved density of states (DOS) of standard Kohm-Sham DFT.
Note also that aside from the pure interacting viewpoint, the effect of temperature in DFT+(sic)DMFT is not only to change the occupational features of the spectrum according to Fermi-Dirac statistics as in finite-$T$ DFT. Here, the temperature introduces also a coherence scale for the electronic excitations, above which those cease to exist.

\begin{figure*}[t]
      \includegraphics[width=\linewidth]{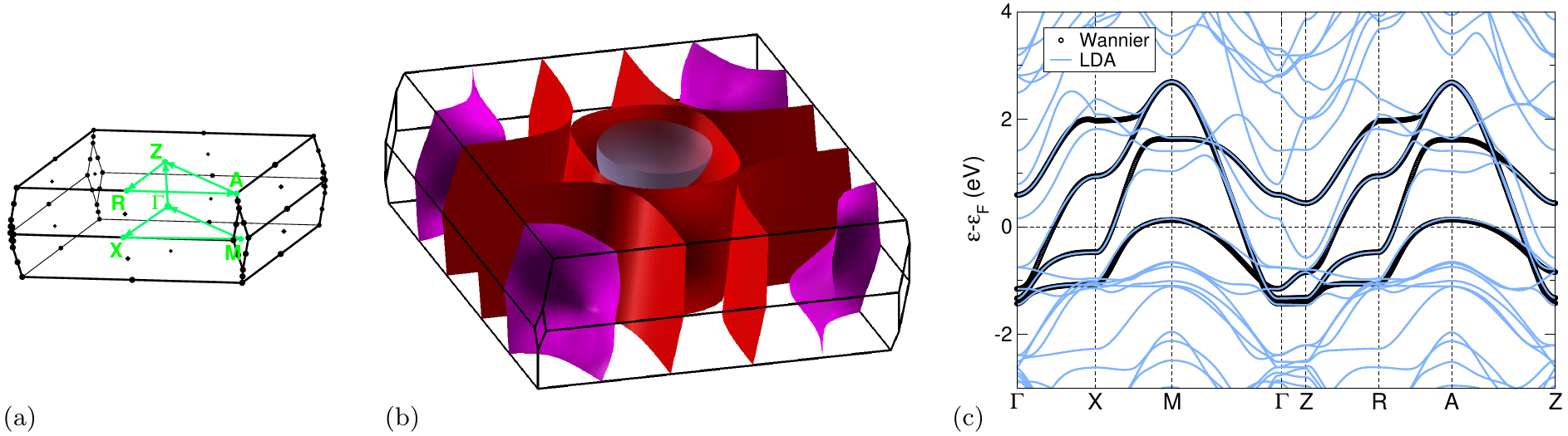}
\caption{Additional DFT data for the $Fmmm$ phase. (a) BZ with high-symmetry lines. (b) Fermi surface with $\alpha,\beta$ sheets (red), $\gamma$ sheet (magenta) and the La-based self-doping sheet (grey transparent). (b) Complete bandstructure (lightblue) and downfolded four-band Ni-$e_g$ based maximally-localized Wannier dispersion (black) for the low-energy region.} \label{SMfiglda}
\end{figure*}

The maximally-localized Wannier construction was performed by initial projection onto the Ni-$e_g$ orbitals in the $Fmmm$ primitive cell. The comparison of the downfolded low-energy Wannier bands with full LDA bands is depicted in Fig.~\ref{SMfiglda}c along with the original LDA Fermi surface (FS) in
Fig.~\ref{SMfiglda}b (with an illustration of the high-symmetry lines in the Brillouin zone (BZ) in Fig.~\ref{SMfiglda}a). Note that the La-based self-doping band forms electron pockets around the $Z$ point. Those pockets are excluded from the Wannier construction, which is justified by their disappearance in the interacting regime described by the comprehensive DFT+sicDMFT approach.

\subsection{Solution of the linearized gap equations}
\subsubsection{Derivation of equation}
The multiorbital random phase approximation (RPA) allows to derive the linearized gap equation as first done in Ref.~\cite{graser2009near}. In the following we will restrict ourselves to the paramagnetic state. The first step is to calculate the multiorbital charge and spin susceptibilities which go into the pairing vertex. The Matsubara formalism allows writing the bare susceptibility in terms of non-interacting Green's functions
\begin{equation}
G_{l_1,l_2}(\mathbf{k},i\omega_n) = \sum_{\mu}  \frac{a_{l_1}(\mu,\mathbf{k}) a_{l_2}^*(\mu,\mathbf{k})}{i\omega_n - \epsilon_{\mu\mathbf{k}}},
\label{Eq:GF}
\end{equation}
where $a_{l}(\mu,\mathbf{k})$ denotes the matrix elements of the unitary transformation from the orbital to band space of the non-interacting part of the Hamiltonian given in Eq.~1 of the main text. It connects the l-th orbital with the $\mu$-th band $\epsilon_\mu$. After summing over the Matsubara frequencies, the analytical continuation of the bare susceptibility follows as:
\begin{widetext}
\begin{equation}
(\chi_0)_{l_1l_4}^{l_2l_3}(\mathbf{q},\omega+i\delta)
=  \frac{1}{N}  \sum_{\mathbf{k}} \sum_{\mu\nu} 
a_{l_1}(\mu,\mathbf{k}+\mathbf{q}) a^{*}_{l_2}(\mu,\mathbf{k}+\mathbf{q}){a}_{l_3}(\nu,\mathbf{k}) {a}^{*}_{l_4}(\nu,\mathbf{k}) \frac{f(\epsilon_{\nu\mathbf{k}})-f(\epsilon_{\mu\mathbf{k}+\mathbf{q}})}{(\epsilon_{\mu\mathbf{k}+\mathbf{q}}-\epsilon_{\nu\mathbf{k}} )-\omega-i\delta}
\label{Eq:BareSus}.
\end{equation}
\end{widetext}
This expression corresponds to a bare bubble diagram with the upper (lower) pair of indices being at the left (right) vertex.
\newpage

\begin{figure*}[t]
      \includegraphics[width=\linewidth]{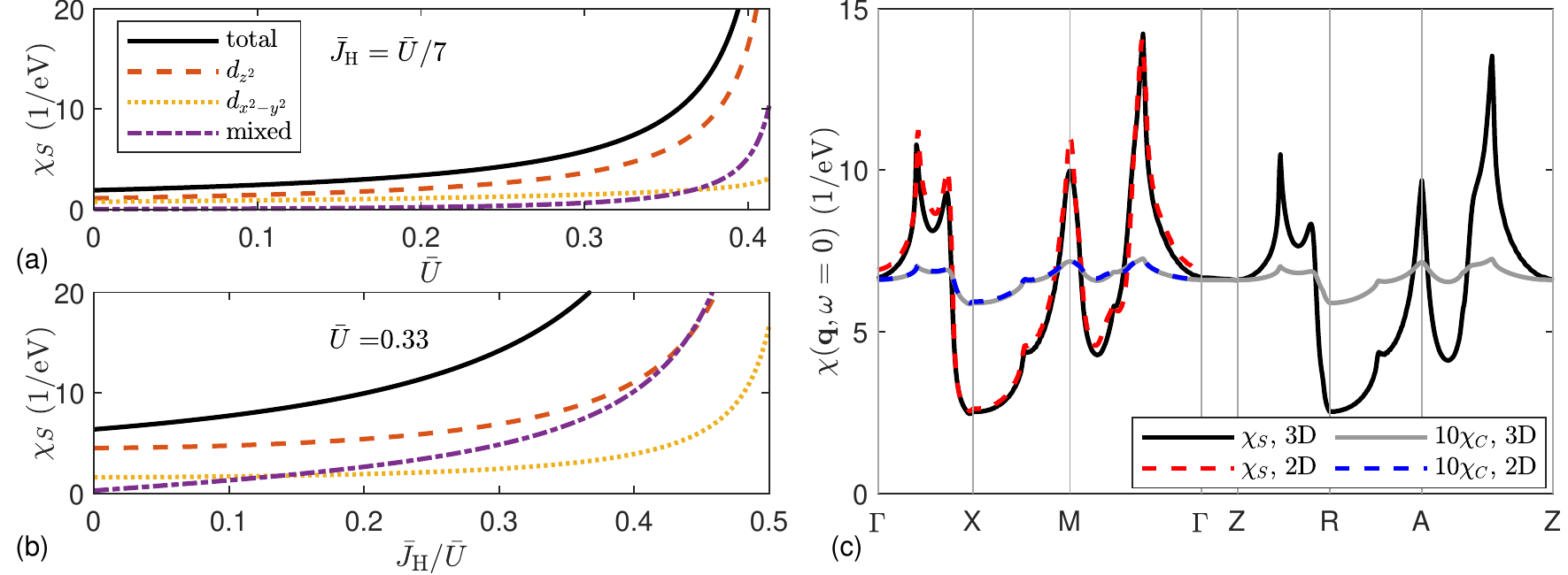}
\caption{RPA susceptibility $(\chi(\mathbf{q},\omega = 0))$ calculated at $T=80$\,K. (a) Spin susceptibility and its orbital resolved contributions calculated at $\mathbf{q}/\pi \approx (0.45,0.45)$ for $\bar{U} = 0.36$ and $\bar{J}_{\rm H} = \bar{U}/7$. (b) As in (a) but as a function of $\bar{J}_{\rm H}$ for constant $\bar{U}=0.33$.  (c) Comparison of spin and charge susceptibilities $\chi_S$ and $\chi_C$ calculated from two-dimensional and three-dimensional approaches. Note that the charge susceptibility is multiplied by a factor of 10.} \label{SMfig2}
\end{figure*}

The on-site interaction part of the Hamiltonian includes intraorbital (interorbital) Hubbard repulsion $\bar{U}$ ($\bar{U}'$), Hund's coupling $\bar{J}_{\rm H}$ and pair-hopping $\bar{J}'$:
\begin{widetext}
\begin{align}
	H^{(i)}_{int} = 
	\frac{\bar{U}}{2} \sum_{il\sigma} 
	c^{\dagger}_{il\sigma}c^{\dagger}_{il\bar{\sigma}}
	c_{il\bar{\sigma}}c_{il{\sigma}}
	+ \bar{U}' \sum_{il<l' \sigma\sigma'} 
	c^{\dagger}_{il\sigma}c^{\dagger}_{il'\sigma'}
	c_{il'{\sigma'}}c_{il{\sigma}} +
	 \bar{J}_{\rm H} \sum_{il<l' \sigma\sigma'} 
	c^{\dagger}_{il'\sigma}c^{\dagger}_{il\sigma'}
	c_{i'\sigma'}c_{il{\sigma}}
	+ \frac{\bar{J}'}{2} \sum_{il\neq l'\sigma} 
	c^{\dagger}_{il\sigma}c^{\dagger}_{il\bar{\sigma}}
	c_{il'\bar{\sigma}}c_{il'{\sigma}},
\end{align}
\end{widetext}
with the creation operator $c^{\dagger}_{l\sigma}$ creating an quasiparticle with orbital $l$ and spin $\sigma$ at site $i$. Assuming spin-rotational invariance for the interactions enforces two relations $\bar{U}' = \bar{U} -2\bar{J}_{\rm H}$ and $\bar{J}' = \bar{J}_{\rm H}$. Therefore, the interaction may be parameterized using $\bar{U}$ and $\bar{J}_{\rm H}$, which are in the following given in units of the bandwidth. 
Within RPA the interaction gives various possibilities to connect the bare susceptibility bubbles, which can be expressed algebraically through a matrix multiplication 
\begin{align}
	(\chi_0)_{\Tilde{l}_1\Tilde{l}_4}^{\Tilde{l}_2'\Tilde{l}_3'} (\Tilde{U})^{\Tilde{l}_1'\Tilde{l}_4'}_{\Tilde{l}_2'\Tilde{l}_3'} (\chi_0)_{\Tilde{l}_1'l_4'}^{\Tilde{l}_2\Tilde{l}_3}.
\end{align}
The matrix indices are to be understood as the combination of the two upper (lower) indices. The matrix multiplication is performed by contracting the upper and lower indices from left to right. Here we use $\Tilde{l} = (l,\sigma)$. If $\sigma_1=\sigma_2$ and $\sigma_3=\sigma_4$, then $\chi_0$ is the same as in Eq.~(\ref{Eq:BareSus}), and 0 otherwise. The components of the matrix $\Tilde{U}$ can be determined from the interaction part. This matrix multiplication includes all the first-order diagrams of the expansion and the higher orders are constructed in a similar fashion by appending matrices $\Tilde{U}\chi_0$ to the right. The resulting Dyson equations can be diagonalized in spin space and the two independent blocks correspond to the spin and charge susceptibilities. These susceptibilities are given by
\begin{align}
	\chi_{S/C} &= \left[ 1 \mp \chi_0	U_{S/C} \right]^{-1}	\chi_0.
\end{align}
The physical observables are obtained by contracting incoming and outgoing orbital indices at both vertices, i.e., by contracting $l_1$ with $l_4$ and $l_2$ with $l_3$\cite{graser2009near}. Therefore, the physical susceptibility includes contributions from various orbital combinations. The elements of the on-site interaction matrices in the Dyson equations for orbitals located on the same atom are given by
\begin{widetext}
\begin{align}
(U_C)_{l_2l_3}^{l_1l_4} = 
\begin{cases}
\bar{U}  &l_1=l_2=l_3=l_4  \\
2\bar{J}_{\rm H}-\bar{U}' &l_1=l_2\neq l_3=l_4  \\ 
2\bar{U}'-\bar{J}_{\rm H}  &l_1=l_4\neq l_2=l_3 \\
\bar{J}' &l_1=l_3\neq l_2=l_4 \\
0  & otherwise.  
\end{cases}
\hspace{1cm}
(U_S)_{l_2l_3}^{l_1l_4}  = 
	\begin{cases}
		\bar{U}  &l_1=l_2=l_3=l_4  \\
		\bar{U}' &l_1=l_2\neq l_3=l_4  \\ 
		\bar{J}_{\rm H}  &l_1=l_4\neq l_2=l_3  \\
		\bar{J}' &l_1=l_3\neq l_2=l_4 \\
		0  & otherwise.  
	\end{cases}
\end{align}
\end{widetext}
To ensure that we stay in the paramagnetic regime, we first examined the parameters for which the matrix $(1 - \chi_0 U_{S})$ has zero eigenvalues, indicating an antiferromagnetic instability. The location of the instability has the shape of a curve in the ($\bar{U}$,$\bar{J}_{\rm H}$)-parameter space. We therefore show the spin susceptibility as a function of $\bar{U}$ with fixed $\bar{U}/\bar{J}_{\rm H}$ and as a function of $\bar{J}_{\rm H}/\bar{U}$ for fixed $\bar{U}$ in Fig.~\ref{SMfig2}a and  Fig.~\ref{SMfig2}b at the largest peak, located at $\mathbf{q}/\pi \approx(0.45,0.45)$. Analyzing the orbital content of the susceptibilities we find that the magnetic instability originates from the mixed $d_{z^2}$ and $d_{x^2-y^2}$ as well as pure $d_{z^2}$ orbital contributions. The mixed contributions can be tuned with $\bar{J}_{\rm H}$ becoming dominant near $\bar{J}_{\rm H}=\bar{U}/2$. 

The summation in Eq.~\ref{Eq:BareSus} is performed at constant $k_z$. However, we utilize the first and second BZ for integration. This includes the $k_z = 0$ sheet and the $k_z = 1/2$ sheet of the BZ as follows from the face-centered cubic structure of the lattice. Figure~\ref{SMfig2}c justifies the use of two-dimensional integration by comparing the charge and spin susceptibility with the full three-dimensional integration along the $\Gamma-X-M-\Gamma$ high-symmetry lines. Note also the quantitative similarity to the $Z-R-A-Z$ path.

\begin{figure*}[t]
      \includegraphics[width=\linewidth]{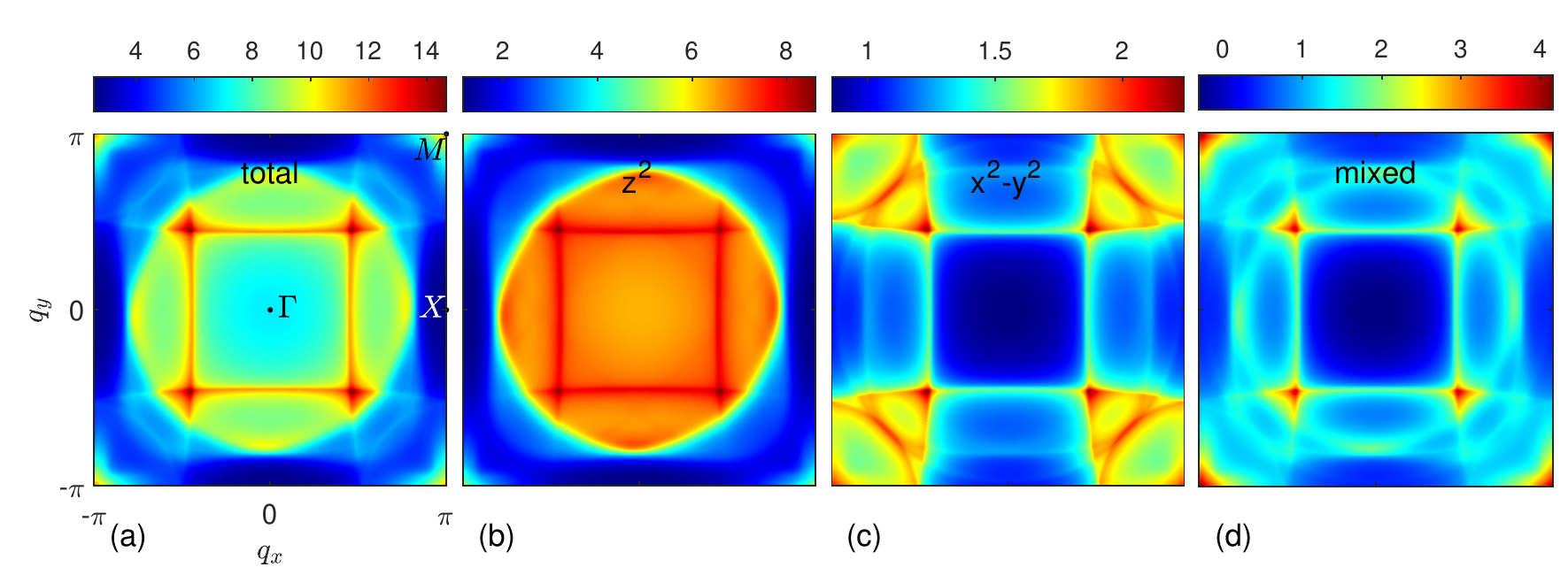}
\caption{RPA spin susceptibility Re$(\chi_S(\mathbf{q},\omega = 0))$ calculated at $T=80$\,K for $\bar{U} = 0.36$ and $\bar{J}_{\rm H} = \bar{U}/7$. (a) Total contribution as in Fig.~4(a) of the main text. (b)-(d) Different orbital resolved contributions to Re$(\chi_S(\mathbf{q},\omega = 0))$. Colorbars are in units of 1/eV.} \label{SMfig3}
\end{figure*}

Figure~\ref{SMfig3} shows the orbital-resolved contributions to the spin susceptibility presented in the constant $k_z$-plane. For the chosen values of $\bar{U} = 0.36$ and $\bar{J}_{\rm H} = \bar{U}/7$, the largest contributions originate again from the $d_{z^2}$ orbital.
The dominance of the $d_{z^2}$ orbital over $d_{x^2-y^2}$ is due to the flatness of the $d_{z^2}$ pocket around the $M$ point. 
The shape of the orbital-resolved spin susceptibility agrees with the results obtained in Ref.~\cite{luo23}.

Within RPA approach the Cooper-pairing vertex can be written in terms of the charge and spin susceptibilities for the paramagnetic case \cite{graser2009near}. The fully momentum-symmetrized interaction vertex projected to the spin singlet channel reads
\begin{widetext}
\begin{equation}
    (\Gamma_\mathrm{s})_{l_1 l_4}^{l_2 l_3}(\mathbf{k},\mathbf{k'}) = \left[ \frac{3}{2} U_S \frac{\chi_S(\mathbf{k}+\mathbf{k'}) + \chi_S(\mathbf{k}-\mathbf{k'})}{2} U_S - \frac{1}{2} U_C \frac{\chi_C(\mathbf{k}+\mathbf{k'}) + \chi_C(\mathbf{k}-\mathbf{k'})}{2} U_C + \frac{1}{2}U_S + \frac{1}{2}U_C \right]_{l_1 l_4}^{l_2 l_3}
\end{equation}
\end{widetext}
and is projected back to band space via unitary transformation to
\begin{widetext}
\begin{equation}
    (\Gamma_{s})_{\mu \nu}(\mathbf{k}, \mathbf{k'}) = \sum_{l_1 l_2 l_3 l_4} \mathrm{Re} \left[ a_{l_1}(\nu,\mathbf{k}) a^{*}_{l_2}(\mu,\mathbf{k'}){a}_{l_3}(\nu,-\mathbf{k}) {a}^{*}_{l_4}(\mu,-\mathbf{k'}) (\Gamma_\mathrm{s})_{l_1 l_4}^{l_2 l_3}(\mathbf{k},\mathbf{k'}) \right].
\end{equation}
\end{widetext}
The interaction vertex includes intraband pairing and the interband Cooper-pair scattering. We did not consider the direct interband Cooper-pairing as this would imply pairing with a finite momentum. Furthermore, notice that the overlap of the $\alpha$ and $\beta$ sheets at the FS is not promoting interband pairing in our on-site interaction approach because their orbital weight is linked to the two distinct nickel atoms of the bilayer. The linearized gap equation
\begin{widetext}
\begin{equation}
    \lambda_i^{s}  \Delta_i^{s}(\mu, \mathbf{k}) = \left[-\frac{1}{V_\mathrm{BZ}} \sum_\nu \int_\mathrm{FS_\nu} \mathrm{d}S' \frac{(\Gamma_{s})_{\mu \nu}(\mathbf{k, k'})}{|\mathbf{v}_\mathrm{F}(\nu, \mathbf{k'})|} \right] \Delta_i^{s}(\nu, \mathbf{k'})
    \label{Eq_linGapEq}
\end{equation}
\end{widetext}
is solved in the spin singlet channel. The integration is carried out over $\mathbf{k'}$, where the bands $\nu$ intersect the FS. $V_\mathrm{BZ}$ represents the volume of the BZ and $v_\mathrm{F}(\nu, \mathbf{k'})$ the Fermi velocity at $\mathbf{k'}$ in the $\nu$-th band. The largest eigenvalue $\lambda_i$ corresponds to the leading instability $i$, which has a superconducting gap $\Delta_i(\mu, \mathbf{k})$ as corresponding eigenvalue.

\subsubsection{Dependence of  the superconducting gap structure on the variation of the band structure}

\begin{figure*}[t]
      \includegraphics[width=\linewidth]{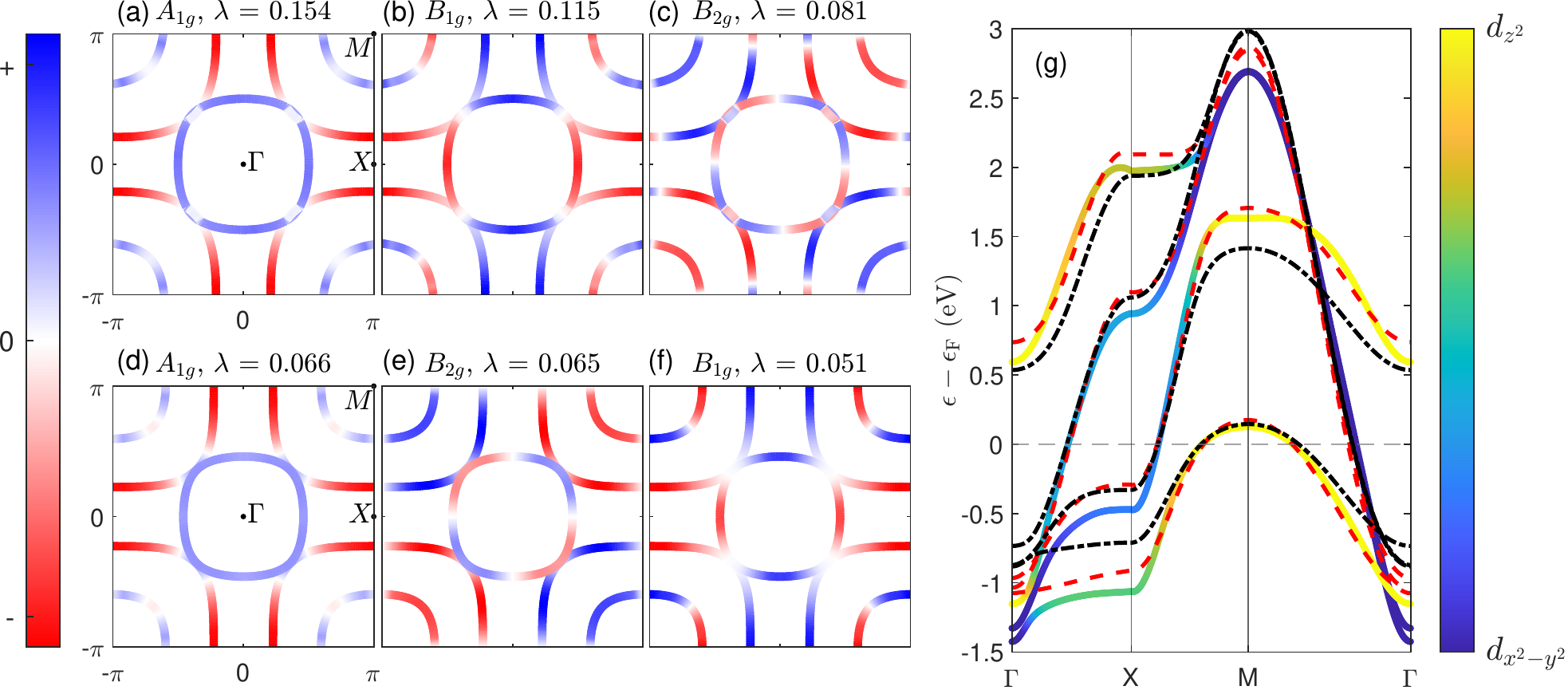}
\caption{Leading and sub-leading solutions to the linearized gap equation using the model from the main text reduced to a tetragonal BZ and next to next nearest neighbor hopping for $k_z=0$ cuts with interactions $\bar{U}=0.36$ and $\bar{J}_{\rm H}=\bar{U}/7$ (a-c) and using the model by Luo {\sl et al.}~\cite{luo23} at $\bar{U}=0.26$ and $\bar{J}_{\rm H}=\bar{U}/7$ (d-f). For all solutions, the corresponding order parameter symmetry and eigenvalue are indicated above. (g) Comparison of the band structures for the full model as in the main text with orbital character of the bands indicated by colorbar, the reduced model (red) and the model by Luo {\sl et al.} (black).} \label{SMfig4}
\end{figure*}

We conducted further research to determine the extent to which the solutions to the linearized gap equation depend on the details of the band structure, including the position of the van Hove singularity at the $X$-point. For this purpose, we also solved Eq.~\ref{Eq_linGapEq} for two simpler tight-binding models. One is the model from the main text reduced to hopping up to the $(i,j,k)$-neighbor with $i,j,k\in\{-1,0,1\}$, compared to $i,j,k\in\{-6,...,6\}$ used in the main text. Furthermore, we are solely considering the $k_z = 0$ layer of the susceptibility. The other tight-binding model is taken from Luo {\sl et al.}~\cite{luo23}, in which only nearest and next-nearest hopping integrals are included and a tetragonal BZ is used to approximately describe the system. The band structures of the models are compared in Fig.~\ref{SMfig4}g. The most striking difference near the FS is the position of the van-Hove singularity at the $X$-point of the $d_{x^2-y^2}$-orbital, which leads to a reduced Fermi velocity for the $\beta$-sheet in the model of Luo {\sl et al.}. Indeed, for large values of $\bar{U}$, spin fluctuations, originating from the scattering between the $\beta$- and $\gamma$-sheets overshoots the intraband scattering within the $\gamma$-sheet. As a consequence, the $A_{1g}$($s_{\pm}$)-wave solution become increasingly predominant. In contrast, the intraband spin fluctuations within the $\gamma$-sheet remains prevalent in our model presented in the main text, preventing the emergence of an $s_{\pm}$-wave solution in that model. Additionally, spin fluctuations contributions stemming from scattering between $\gamma$ and $\alpha$ bands are also stronger than those from $\gamma$ and $\beta$ interband scattering for the model presented in the main text.

\begin{figure*}[t]
      \includegraphics[width=\linewidth]{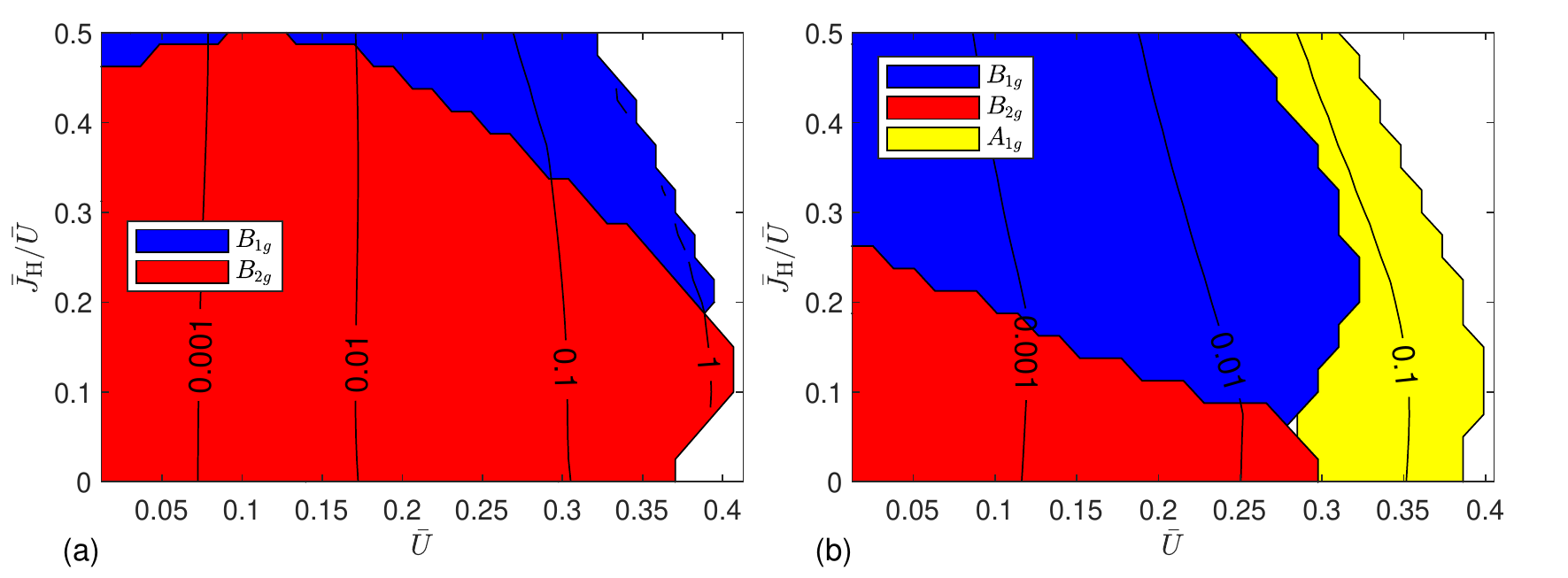}
\caption{Phase diagram of the leading solution of the linearized gap equation for the model presented in the main text (a) and the reduced model (b).} \label{SMfig5}
\end{figure*}

The three leading solutions of the linearized gap equation for our reduced model and the model by Luo {\sl et al.} are shown in Fig.~\ref{SMfig4}a-c and Fig.~\ref{SMfig4}d-f, respectively, for which we choose $\bar{J}_{\rm H}=\bar{U}/7$ and $\bar{U}$ close to the magnetic instability. Note that the magnetic instability is also found at higher $\bar{U}$ values for our description compared to Luo {\sl et al}. For these interaction values, the leading solution is a $A_{1g}$ ($s_{\pm}$-wave) and the $B_{1g}$ ($d_{x^2-y^2}$-wave) and $B_{2g}$ ($d_{xy}$-wave) solutions are only sub-leading in both cases. In this regime the eigenvalues are of comparable size. The still sizeable spin fluctuation contribution from the scattering within the $\gamma$-sheet enforces a suppression of the gap at the diagonal of the $\gamma$-sheet in Fig.~\ref{SMfig4}a and a sign change in Fig.~\ref{SMfig4}d at the same location. These features vanish if the scattering between $\beta$ and $\gamma$-sheets is further increased by e.g. moving the van Hove singularity further towards the FS. For smaller $\bar{U}$, i.e. further away from the magnetic instability, the $B_{1g}$ and $B_{2g}$ solutions become leading for large and small Hund's coupling $\bar{J}_{\rm H}$, respectively, similar to our model presented in the main text. The corresponding ($\bar{U}$ and $\bar{J}_\mathrm{H}$)-phase diagram is displayed in Fig.~\ref{SMfig5}. In Fig.~\ref{SMfig5}a and Fig.~\ref{SMfig5}b the model presented in the main text and our reduced model are compared, respectively. For the tight-binding model by Luo {\sl et al.} we find a similar phase diagram as in Fig.~\ref{SMfig5}b. Comparing the phase diagrams, the importance of the interplay between the VHS of the $d_{x^2-y^2}$-orbital and the flatness of the $d_{z^2}$-orbital mediated through long-range hopping, becomes evident. 

Finally, it should be noted that in the model presented in the main text, the most dominant scattering vectors are not only limited to the first BZ cut with $k_z = 0$. The most pronounced difference is in the scattering between the $\gamma$- and $\alpha$-sheets, where the scattering amplitude of the $\alpha$-sheet in the $k_z = 1/2$ layer is up to 50 \% stronger than in the $k_z = 0$ layer. To understand this difference, note first the anisotropy in the inverse of the Fermi velocity in Fig.4b of the main text peaks in the $k_z = 1/2$ layer, and second that the FS of both the $\alpha$- and $\gamma$-sheets are not perfectly two-dimensional. The bulbous $\alpha$-cylinder is more edgy and has a smaller cross-section in the $k_z = 1/2$ layer. Therefore, we conclude that the approximation of the system to a purely tetragonal BZ might be insufficient to capture all the important details relevant for the scattering process and details of the spin fluctuations.

\subsubsection{Numerical details}
The calculation of the spin susceptibility and the FS involves taking a cut through the first and second BZ to account for scatterings between sheets at $k_z = 0$ and $k_z = 1/2$, as justified above. The integral in Eq.~\ref{Eq_linGapEq} is transferred to two dimensions, accordingly.

The spin susceptibility utilized for interpolating the susceptibility in the linearized gap equation, as demonstrated in Fig.~4a of the main text along with Fig. \ref{SMfig3} of the supplementary, is computed on a $256^2$ $\mathbf{q}$-mesh according to Eq.~\ref{Eq:BareSus}. The $\mathbf{k}$ integral is carried out with a resolution of $512^2$ $\mathbf{k}$-points. In the cut along the high-symmetry lines shown in Fig.~\ref{SMfig2}, the same integral resolution is used for the 2D calculation and $256\times256\times128$ $\mathbf{k}$-points for the 3D case. We use 300 points for each segment of the path. In Eq.~\ref{Eq:BareSus} we use $\delta = 10^{-3}$ meV. 

The FS integral in Eq.~\ref{Eq_linGapEq} is evaluated by parametrizing the FS with a set of points at which the Fermi velocity and transformation matrices are calculated. The bare susceptibilities $\chi_0(\mathbf{k}-\mathbf{k}')$ and $\chi_0(\mathbf{k}+\mathbf{k}')$ are derived through linear interpolation of the fine susceptibility grid used to construct Fig.~\ref{SMfig3}.

Such parametrisation can be obtained by a standard contour algorithm, which is done for a fine momentum mesh with $1024^2$ points. As many of these points are close to each other and the computational cost increases with the number of points, we want the points used to be equidistantly spaced. To accomplish this, we employ the ``interparc'' algorithm \cite{interparc}, which uses the initial points to find differential equations that accurately describe the parameterized curve. A solver for ordinary differential equations enables one to interpolate equidistant points. For the creation of Fig.~4c and Fig.~4d in the main text, the parameterization contained 1368 points.


\end{document}